\documentstyle[12pt,axodraw]{article}

\begin{document}
\newcommand{\be}{\begin{equation}}
\newcommand{\ee}{\end{equation}}
\newcommand{\al}{\alpha}
\title
{
\vspace*{-40mm}
\bf
\begin{flushright}
{\large  TTP/99-28 \\[-2mm]
June 1999\\[15mm]}
\end{flushright}
\bf
Sudakov logarithms in electroweak processes
}
\author{
 J.H.K\"uhn\\[-1mm] 
  {\small {\em Institut f{\"u}r Theoretische Teilchenphysik,
  Universit{\"a}t Karlsruhe}}\\[-2mm]
  {\small {\em D-76128 Karlsruhe, Germany}}\\
  and\\
  A.A.Penin\thanks{On leave from  Institute for Nuclear 
  Research of Russian Academy   of Sciences}\\[-2mm]
  {\small {\em II. Institut f{\"u}r Theoretische Physik,
  Universit{\"a}t Hamburg}}\\[-2mm]
  {\small {\em  22761 Hamburg, Germany}}
}
\date{}

\maketitle

\begin{abstract}
The  dominant electroweak double logarithmic  corrections 
to the process $e^+e^-\rightarrow f\bar f$ at high energy
are found in all orders of perturbation theory.
In contrast to results in simple Yang-Mills theories
these corrections do not exponentiate.
\\[2mm]
PACS numbers: 12.15.Lk 
\end{abstract}

\thispagestyle{empty}

\newpage
The double logarithmic ``Sudakov'' corrections in Abelian \cite{Sud} 
and non-Abelian \cite{CorTik} gauge theories
are essentially determined  by the infrared structure of the theory.
They dominate in the limit of fixed-angle scattering when 
all the  invariant energy and momentum transfers
of the process become much larger than the 
typical mass scale of the particles running inside the loop.

In electroweak processes at energies 
far larger than $M_Z$ or $M_W$
the corresponding corrections due to the virtual 
$Z$ $(W)$ boson exchange scale like a power of 
${g^2\over 16\pi^2}\log^2\left({s\over M_{Z,W}^2}\right)$ 
\cite{KDB}\footnote{For the massless photon the corrections are 
infrared divergent and one has to take into account the real 
soft photon radiation to obtain an infrared safe cross section.
We do not consider this effect and focus on the pure electroweak 
corrections.}. They grow rapidly with energy and become
dominant  in the TeV region available at the 
LHC or the Next Linear Collider. 

In this paper we consider the dominant  electroweak 
corrections to the process $e^+e^-\rightarrow f\bar f$ at high
energy.  In fact, the results are equally well applicable to any process
$f' \bar f' \to f \bar f$ and we will therefore keep the discussion
quite general throughout this paper.
The one loop double logarithmic corrections to this process
reach $\sim -10\%$ at  $s\sim 1~{\rm TeV}^2$ \cite{CiaCom}.
Similarly large negative corrections were observed in \cite{Hollik} for
quark pair production in hadronic collisions.
The two loop  terms may well be  
comparable  to the subleading  one loop  
corrections \cite{Hol}  and therefore
have to be taken into account to guaranty  reasonable accuracy of the
theoretical predictions.  Though only the two loop corrections
are of practical interest we give the general result  for
the double logarithmic contribution 
in an arbitrary order of perturbation theory. 
 
In  Born approximation the chiral amplitudes
of the process $f'\bar f'\rightarrow f\bar f$ can be written in  the form 
\be
M_{IJ}^{B}={ig^2\over s}\left(T^3_fT^3_{f'}+
t^2_W{Y_fY_{f'}\over 4}\right)
\label{amplit0}
\ee
where ${I,~J}={L,~R}$, $t_W=\tan{\theta_W}$, 
$T^3_f$ is  the isospin and $Y_f$ is the hypercharge 
of the fermion.  

We study the fixed-angle regime and restrict the analysis to the 
double logarithmic contributions  
which originate from the interplay between the soft and 
collinear  singularities. All  single logarithmic contributions 
from soft, collinear or ultraviolet divergencies are neglected. 

The structure of the electroweak corrections  depends 
crucially on  the chirality of the fermions. 
Let us start with the non-Abelian interaction of left-handed
fermions.  It has been demonstrated \cite{CorTik,FreTay}
that the leading infrared divergencies exponentiate for
Yang-Mills theories and one obtains for each on-shell fermion line the
suppression factor
\be
\exp{\left(-{C_F\over 2}L(s)\right)}
\label{exp}
\ee
where 
\be
L(s)={g^2\over 16\pi^2}\log^2\left({s\over M^2}\right),
\label{log}
\ee
$C_F$ is the  Casimir operator ($C_F=3/4$ for $SU(2)$) 
and $M$ is the infrared regulator (the gauge boson mass).
The external fermions are assumed to be in the fundamental
representation. 
The fact that the double logarithmic corrections are determined
by the external on-shell  particles and do not depend on the
details of the process   has been proven in ref.~\cite{FreTay}
where the authors demonstrate that, in a physical (Coulomb or axial) 
gauge, the  double logarithms originate from the self energy corrections
to the  external lines. 

However, this result  cannot  
be directly applied to the electroweak case because the $Z$ boson, 
the mass eigenstate, is a mixture of $A^3_\mu$,
the pure $SU(2)_L$, 
and $B_\mu$, the Abelian hypercharge  components. Hence,  
it is the difference of  the coupling
of  $Z$ and $W$ bosons  to fermions, that 
renders eq.~(\ref{exp}) invalid. Nevertheless,
it is quite straightforward to  obtain the one loop result
for the case of interest using this equation.      
The  lowest order double logarithmic contribution in the Coulomb (axial)
gauge is determined by the diagram displayed in Fig.~1 \cite{FreTay,FreMeu} 
and in the absence of mixing 
$(\theta_W=0)$ the correction reads 
\be
-{C_Fg^2\over 32\pi^2}\log^2\left({s\over M^2}\right)
\label{term}
\ee   
which is the first non-trivial term in the expansion of eq.~(\ref{exp})
in $g^2$. The  neutral Yang-Mills boson $A^3_\mu$ 
gives one third of this term. 
Due to the trivial structure of the first order correction
one obtains the contribution of the $Z$  boson 
by substituting the $A^3_\mu$ coupling $gT^3_f$
by the $Z$ boson coupling $g(T^3_f-s^2_WQ_f)/c_W$ in the 
expression for the $A^3_\mu$  contribution. Here 
$s_W=\sin{\theta_W},~c_W=\cos{\theta_W}$  and $Q_f$ is 
the electric charge of the fermion.
The contribution of the $W$ boson amounts to  two third of eq.~(\ref{term}). 
Thus in lowest order the double logarithmic corrections
give the factor $1-F^f_L$ for each pair of the 
incoming/outgoing left-handed fermion lines where
\be
 F^f_L=\left({1\over 2}+{1\over\ 4c^2_W}
+t^2_W\left(s^2_WQ^2_f-2T^3_fQ_f\right)\right)L(s).
\label{factorl}
\ee 

The case of the right-handed fermions is rather simple
since their  interaction to  $Z$ boson is Abelian.
Therefore the  double logarithms
exponentiate to 
\be
\exp{\left(-{t^2_Ws^2_WQ^2_f\over 2}L(s)\right)}
\label{expright}
\ee
for each on-shell fermion leg.
In  first order this  gives the factor $1-F^f_R$ 
for each pair of the  incoming/outgoing right-handed 
fermion lines where
\be
 F^f_R=t^2_Ws^2_WQ^2_fL(s).
\label{factorr}
\ee
Thus the first order  amplitudes  in the 
double logarithm approximation take the form
\be
M_{IJ}^{(1)}=M_{IJ}^{B}\left(1-F^f_I-F^{f'}_J\right).
\label{amplit1}
\ee
This result coincides with the more cumbersome expression 
obtained in ref.~\cite{CiaCom} by direct evaluating the Feynman
diagrams in the eikonal approximation. 
Note that in  \cite{CiaCom} the covariant gauge 
was used for the calculation and the double logarithms 
originate from both  vertex and  box diagrams.    
Anyhow, one loop  corrections are
known exactly \cite{Hol}
so the first order  result in  double logarithmic approximation
is of no specific interest.

Let us now turn to the two loop analysis. 
 For the right-handed fermions the correction is determined by the 
second term of the expansion of eq.~(\ref{expright})
in $g^2$.  For the left-handed fermions the situation is 
less trivial. In this order the contribution from the $Z$ boson 
cannot be obtained by changing the coupling in the contribution
from the $A^3_\mu$ boson to   eq.~(\ref{exp}) since exponentiation 
breaks down explicitly. To see this let us consider 
the two ``rainbow'' diagrams  in Fig.~2 which give  part of the 
second order double logarithmic contribution  \cite{FK}.
The rest is determined by similar
diagrams with two $Z$ or two $W$ bosons in the loops. 
In the absence of mixing  diagrams $(a)$     
and $(b)$ give the same result. 
 For non-zero $\theta_W$ this is not true due to 
the presence of the electric charge in 
the $Z$ boson coupling to fermions. Moreover, the  
result for the diagram $(a)$
can be obtained from the corresponding $\theta_W=0$ diagram 
by  the same change of the $Z$ boson  coupling   
$gT^3_f\rightarrow g(T^3_f-s^2_WQ_f)/c_W$ 
as  used in the analysis of the first order corrections
while the diagram $(b)$  cannot and therefore is non-exponential.
In fact,   diagram $(b)$  is the only source of the
non-exponential corrections in the second order. 
Writing the sum $(a)+(b)$ as  $2\cdot(a)+((b)-(a))$ one finds that
$2\cdot(a)$ contribution corresponds to 
a part of the second term in the expansion of the exponent
\[
1-F^f_L/2+{{(F^f_L/2)}^2\over 2!}+\ldots.
\]  
In this way we find that 
for each pair of the  incoming/outgoing left-handed fermion lines 
the leading two loop correction reads:
\be
{1\over 2!}{F^f_L}^2+\Delta_f
\label{secondterm}
\ee
where the first term arises from the expansion of the exponent
and  
\be
\Delta_f={t^2_W\over 8}\left(\left(s^2_WQ^2_{\tilde f}-
2T^3_{\tilde f}Q_{\tilde f}\right)- \left(s^2_WQ^2_f-
2T^3_fQ_f\right)\right)
\label{delta}
\ee
is twice the difference $(b)-(a)$.
Thus, in two loop approximation the leading logarithmic
approximations of the chiral amplitudes read as follows:
\be
M_{IJ}^{(2)}=M_{IJ}^{B}\left(1-\left(F^f_I+F^{f'}_J\right)+
{1\over 2!}\left(F^f_I+F^{f'}_J\right)^2+
\delta_{IL}\Delta_f+\delta_{JL}\Delta_{f'}\right)
\label{amplit2}
\ee
where
\[
\delta_{JL}=\left\{
\begin{array}{l}
0,\qquad J=R,\\
1,\qquad J=L\\
\end{array}\right.
\]
Now it is straitforward to obtain the general double logarithmic 
contribution from the corresponding  ``rainbow'' diagrams. 
In $n$th order instead of the term  
\be
{(-1)^n\over n!}\left({C_F\over 2}L(s)\right)^n
\label{naive}
\ee
of the  expansion of the exponent~(\ref{exp})
we get
\be
{(-1)^n\over n!}\left({C_Fx_fL(s)\over 6}\right)^n\left(1+\sum_{m=1}^n
\sum_{k=0}^{n-m}
2^m
C_{n-m+l-k}^l
C_{m-l-1+k}^{k}
{x_{\tilde f}^k\over x_f^{k+m}}
\right) 
\label{general}
\ee 
where $C_i^j$ are binomial coefficients,  
$l$ is the integer part of $m/2$ and
\[
x_f=4\left({T^3_f-s^2_WQ_f\over c_W}\right)^2.
\]
 For $\theta_W=0$ $(x_f=x_{\tilde f}=1/4)$ eq.~(\ref{general})
is reduced to  eq.~(\ref{naive}).
Note that the non-exponential contribution is suppressed
by the small quantity $s_W^2\sim 0.23$ which can be considered as 
an additional expansion parameter. 
This would essentially simplify
the analysis of the subleading logarithms because 
in the leading order in $s_W^2$  the results for the simple Yang-Mills 
theories \cite{Sen} are applicable.     
The exponential form of the corrections can be restored 
by adding a (hypothetical) contribution of the heavy photon
of mass $M$ (we leave aside the Higgs mechanism 
since the problem of mass generation
is irrelevant for the analysis of the infrared
properties of the theory). In this case 
one would find
\be
\exp{\left(-{C_F+t^2_WY^2_f/4\over 2}L(s)\right)}
\label{exptot}
\ee
instead of   eq.~(\ref{exp}).  The exponentiation holds also 
for the massless photon if considered separately. 
So the reason of the absence of
exponentiation for the  electroweak
double logarithmic  contributions is twofold:\\
i)  The mass eigenstates, namely 
photon and $Z$ boson, have no
definite gauge transformation properties. \\
ii) The photon is massless and has to be treated separately
with real radiation being taken into account, e.g. in a completely
inclusive manner.

The chiral amplitudes determine the differential cross section
of the process 
\[
{d\sigma\over d\Omega}={N_c^{(f)}s\over (16\pi)^2}\left(\left(|M_{LL}|^2+
|M_{RR}|^2\right)(1+\cos{\theta})^2+\left(|M_{LR}|^2+
|M_{RL}|^2\right)(1-\cos{\theta})^2\right)
\]
where $N_c^{(f)}$ is $3$ for quarks and  $1$ for leptons.
With the expression for the  differential cross section
at hands we can compute the leading logarithmic corrections
to the basic observables for $e^+e^-\rightarrow f \bar f$.

In the fixed-angle regime 
and in the leading logarithmic approximation,
the invariants $s$ and $t$ are not distinguished
since the difference is subleading. Therefore we can
integrate the  differential cross section 
over all angles to find the leading result for the total
cross sections and for asymmetries, though formally the double logarithmic
approximation is not valid for the small angles 
$\theta <M/\sqrt{s}$.   
These   corrections 
also do not depend on the choice of  mass, used 
as the scale $M$. Since the difference 
between $M_W$ and $M_Z$ is subleading,
$M=M_W$ is used throughout. In the two loop approximation we find:
\[
\sigma/\sigma_B(e^+e^-\rightarrow Q\bar Q)
=1-1.659L(s)+1.992L^2(s)
\]
\be
\sigma/\sigma_B(e^+e^-\rightarrow q\bar q)
=1-2.173L(s)+2.826L^2(s)
\label{corrsig}
\ee
\[
\sigma/\sigma_B(e^+e^-\rightarrow \mu^+\mu^-)
=1-1.392L(s)+1.495L^2(s)
\]
where $Q=u,c,t$,  $q=d,s,b$. 
Numerically $L(s)=0.07$ and $0.11$
for $\sqrt{s}=1$~TeV and $2$~TeV respectively. Here $M=M_W$
has been choosen for the infrared cutoff and $g^2/16\pi^2=2.7\cdot 10^{-3}$
for the $SU(2)$ coupling evaluated at $\sqrt{s}=1$ TeV.
Clearly, for energies at 1 and 2 TeV
the two loop corrections are huge and amount up to $1\%$ and $4\%$ 
respectively.

 For the  forward-backward asymmetry (the difference of 
the cross section averaged over forward and backward semispheres 
in respect to the electron beam direction
divided by the total cross section) we get
\[
A_{FB}/A_{FB}^B(e^+e^-\rightarrow Q\bar Q)
=1-0.090L(s)+0.120L^2(s)
\]
\be
A_{FB}/A_{FB}^B(e^+e^-\rightarrow q\bar q)
=1-0.140L(s)+0.024L^2(s)
\label{corrafb}
\ee
\[
A_{FB}/A_{FB}^B(e^+e^-\rightarrow \mu^+\mu^-)
=1-0.039L(s)+0.281L^2(s)
\]
 For the left-right asymmetry (the difference of the cross sections 
of the left and right particles production divided by the 
total cross section) we obtain
\[
A_{LR}/A_{LR}^B(e^+e^-\rightarrow Q\bar Q)
=1-1.129L(s)+0.821L^2(s)
\]
\be
A_{LR}/A_{LR}^B(e^+e^-\rightarrow q\bar q)
=1-4.551L(s)+1.123L^2(s)
\label{corralr}
\ee
\[
A_{LR}/A_{LR}^B(e^+e^-\rightarrow \mu^+\mu^-)
=1-13.744L(s)+0.399L^2(s)
\]

 For the loops with a  top quark
running inside one may replace  the  square of logarithm 
in eq.~(\ref{log}) by
\be
\log^2\left({s\over m_t^2}\right)+4\log\left({s\over m_t^2}\right)
\log\left({m_t\over M}\right),
\label{logtop}
\ee
a form  valid with logarithmic accuracy for $m_t>> M$.
Clearly, the difference between logarithm factors in 
eq.~(\ref{log}) and eq.~(\ref{logtop}) is  also subleading. 
Numerically various definitions of $L(s)$ differ by
$\sim 10\%$  at  $s\sim 1~{\rm TeV}^2$ {\it i.e.} the uncertainty 
is of the  order of the generic non-enhanced electroweak
corrections in each order. For the physical applications
this  is important in one loop approximation. 
 Fortunately  the first order corrections are known exactly 
beyond the  double logarithmic approximation. 
At the same time this difference  is small in the two loop
order double logarithmic corrections
which  are of the main interest because they are 
supposed to dominate  the (still unknown) total two-loop  
electroweak corrections.  

To conclude, we have 
found  the dominant double logarithmic electroweak  Sudakov  corrections 
to the process $e^+e^-\rightarrow f\bar f$ at high energy
in all orders of perturbation theory. 
In contrast to results in simple Yang-Mills theories
these corrections do not exponentiate. The explicit expression for the
two loop corrections to the total cross sections and asymmetries
has been obtained. These corrections  reach a few percents size 
in the TeV region  and are crucial for high precision tests of 
the electroweak model at future colliders.

\vspace{5mm}
\noindent
{\large \bf Acknowledgements}\\[2mm]
A.A.Penin gratefully acknowledges discussions
with K.Melnikov.
This work is partially supported 
by Volkswagen Foundation under contract
No.~I/73611, by BMBF under grant number 
BMBF-057KA92P and by the DFG-Forschergruppe 
``Quantenfeldtheorie, Computeralgebra und
Monte-Carlo-Simulation''. 
The work of A.A.Penin is supported in part  by
the Russian Fund for Basic Research under contract
97-02-17065 and Russian Academy of Sciences under contract N37.

\vspace{5mm}
 
\section*{Figure captions}

\noindent
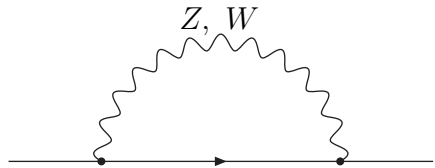
{\bf Fig. 1.} Self-energy correction determining the 
first order double logarithm contribution in 
the Coulomb or axial gauge.

\noindent
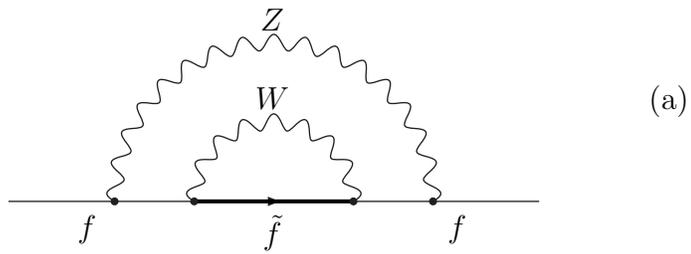
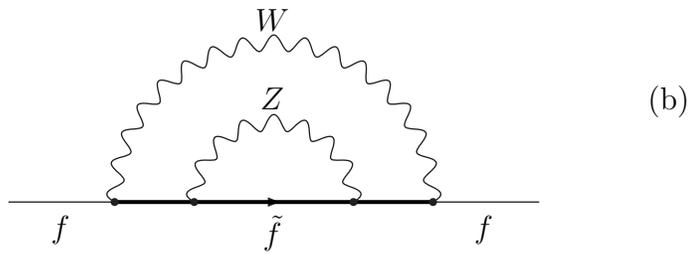
{\bf Fig. 2.} ``Rainbow'' diagrams  giving a part of the 
second order double logarithm contribution. The bold
line corresponds to the fermion isospin partner $\tilde f$.
The fact that the diagram $(b)$ is not equal to the diagram $(a)$
destroys exponentiation.

\newpage

\vspace*{15mm}

\begin{center}\begin{picture}(300,56)(0,0)
\ArrowLine(70,10)(230,10)
\PhotonArc(150,10)(45,0,180){3}{12.5}      
\Vertex(195,10){1.5} \Vertex(105,10){1.5}
\Text(150,60)[cb]{$Z,~W$}
\end{picture} \end{center}

\vspace{5mm}

\begin{center}
{\Large Fig. 1}
\end{center}

\vspace{20mm}

\begin{center}\begin{picture}(300,56)(0,0)
\ArrowLine(50,10)(250,10)
\GBox(120,9.5)(180,10.5){0}
\PhotonArc(150,10)(60,0,180){3}{16.5}   
\PhotonArc(150,10)(30,0,180){3}{8.5}   
\Vertex(180,10){1.5} \Vertex(120,10){1.5}
\Vertex(210,10){1.5} \Vertex(90,10){1.5}
\Text(150,45)[cb]{$W$}
\Text(300,45)[cb]{(a)}
\Text(150,75)[cb]{$Z$}
\Text(80,5)[ct]{$f$}
\Text(220,5)[ct]{$f$}
\Text(150,5)[ct]{$\tilde f$}
\end{picture} \end{center}

\vspace{10mm}

\begin{center}\begin{picture}(300,56)(0,0)
\ArrowLine(50,10)(250,10)
\GBox(90,9.5)(210,10.5){0}
\PhotonArc(150,10)(60,0,180){3}{16.5}   
\PhotonArc(150,10)(30,0,180){3}{8.5}   
\Vertex(180,10){1.5} \Vertex(120,10){1.5}
\Vertex(210,10){1.5} \Vertex(90,10){1.5}
\Text(150,45)[cb]{$Z$}
\Text(300,45)[cb]{(b)}
\Text(150,75)[cb]{$W$}
\Text(70,5)[ct]{$f$}
\Text(230,5)[ct]{$f$}
\Text(150,5)[ct]{$\tilde  f$}
\end{picture} \end{center}

\vspace{10mm}

\begin{center}
{\Large Fig. 2}
\end{center}

\end{document}